\documentclass[twocolumn,showpacs,showkeys,preprintnumbers,amsmath,amssymb]{revtex4}
\usepackage{dcolumn}
\usepackage{bm}
\newcommand{\be}{\begin{equation}}
\newcommand{\ee}{\end{equation}}
\newcommand{\ba}{\begin{eqnarray}}
\newcommand{\ea}{\end{eqnarray}}
\newcommand{\nl}{\nonumber \\}

\begin{document}

\title{Axial coupling from matching constituent quark model to QCD}

\author{S. S. Afonin}
\affiliation{ V. A. Fock Department of Theoretical Physics, St.
Petersburg State University, 1 ul. Ulyanovskaya, 198504 St.
Petersburg, Russia.}


\begin{abstract}
The axial-vector coupling $g_A$ of a constituent quark is
estimated from matching the constituent quark model to the
operator product expansion in QCD in the limit of large number of
colours under some assumptions. The obtained relation is
$g_A\simeq\sqrt{7/11}\approx0.80$, which is in agreement with the
existing model estimates.
\end{abstract}

\maketitle

The constituent quark model (CQM) of Georgi and
Manohar~\cite{georgi} (often called the chiral quark model) has
experienced a great phenomenological success in description of the
strong interactions at low energies, thus giving rise to various
extensions and new applications (see, e.g.~\cite{andr}). The
underlying philosophy of this model is familiar from different
branches of physics --- the idea of rearrangement of physical
degrees of freedom at certain energy scale. The proposed
scenario assumes that in the energy region between the chiral
symmetry breaking (CSB) scale,
$\Lambda_{\text{CSB}}\simeq1\!\!-\!\!1.2$~GeV, and the confinement
scale, $\Lambda_{\text{QCD}}\simeq100\!-\!300$~MeV, the almost
massless strongly interacting ($\alpha_s(\Lambda_{\text{QCD}})>1$)
quarks entering the QCD Lagrangian are effectively rearranged into
heavy weakly interacting
($\alpha_s(\Lambda_{\text{QCD}})\simeq0.28$) constituent quarks
with the effective mass $m\simeq300\!-\!350$~MeV.
Simultaneously, the interaction of fundamental quarks and gluons
is rearranged below $\Lambda_{\text{CSB}}$ into the interaction of
the constituent quarks with the Goldstone bosons --- the pions ---
associated with the spontaneous CSB and, possibly, with the
low-energy gluons. For instance, the $SU(2)_L$ current
in the effective CQM Lagrangian~\cite{georgi} is
\be
\label{cur}
j_{\mu,L}^{\text{CQM}}=\bar{\psi}\gamma_{\mu}(1-g_A\gamma_5)\vec{\tau}\psi+\text{terms
involving $\pi$},
\ee
while the same current in the QCD Lagrangian above
$\Lambda_{\text{CSB}}$ is
\be
\label{cur2}
j_{\mu,L}^{\text{QCD}}=\bar{\psi}\gamma_{\mu}(1-\gamma_5)\vec{\tau}\psi.
\ee
It should be emphasized that, generally speaking, the quark and
gluon fields in the CQM Lagrangian are not the same as the ones in
the QCD Lagrangian. Neglecting the current quark masses, both
fundamental and effective theories are chirally invariant, but the
chiral symmetry is realized non-linearly in the effective theory
in contradistinction to
the linear realization in the fundamental theory, thus the chiral
symmetry undergoes a sort of "rearrangement" of its realization below
$\Lambda_{\text{CSB}}$ rather than breaking.

In this Brief Report we will concern the axial coupling $g_A$ which is
present in current~\eqref{cur}. This coupling is of high
importance in the phenomenology because the $SU(2)_L$
current~\eqref{cur} couples to the $W$ boson, thus triggering some
semileptonic decays. Making use of non-relativistic quark model
wave functions, $g_A$ can be related with the axial constant $G_A$
which parametrizes the amplitude of the nucleon $\beta$-decay, the
relation is~\cite{georgi}
\be
\label{Ga}
g_A=\frac35G_A.
\ee
Taking the modern experimental value for the axial constant~\cite{pdg},
$G_A\approx1.27$, relation~\eqref{Ga} yields the
estimate $g_A\approx0.76$.

As was noted in~\cite{georgi}, $g_A$ should be calculable from
QCD, although this is a hard non-perturbative calculation and
there is still no idea how to perform it. At present there are
only some estimates based on effective models and on the analog of
the Adler-Weisberger sum rule for quark-pion scattering,
see~\cite{rafael} for a brief review and also~\cite{bron,jacob}.

We will consider a quite different way for addressing the problem.
One can try to probe the vacuum by
the vector (V) and axial-vector (A) currents of the fundamental and
effective theories, the analysis and comparison of the corresponding
responses might lead to definite conclusions. It is rather hard
to garment such a general idea with precise calculations, a kind of
guesswork is inescapable to advance. We will  propose an heuristic
way for the analytical realization of this program, the
way which, albeit qualitative, will result in a numerical estimate
for $g_A$.

Our proposal is based on the expectation that the applicability of
the CQM and that of the perturbative QCD should overlap in some
energy region, {\it i.e.} both theories should give the same result in
that region, thus they can be matched (similar ideas of matching
were exploited in various effective models, see,
{\it e.g}.~\cite{andr,match}). First of all, let us estimate the matching
region. The main difference between the CQM and QCD is induced by
the pions as long as they are quark-antiquark bound states in QCD,
while within the CQM the pions represent fundamental fields.
It is expected (see, {\it e.g.}~\cite{jacob}) that the simple chiral
quark model is only applicable in the low-energy region
$\mu<m_{\rho}$, where $m_{\rho}$ is the $\rho$-meson mass,
$m_{\rho}=775.5$~MeV~\cite{pdg}, whereas in the intermediate
energy region, $m_{\rho}<\mu<\Lambda_{\text{CSB}}$, one should
take into account the higher order derivative terms of the pion
field and, probably, the $\rho$ and $\sigma$ mesons and long-range
gluons as explicit degrees of freedom. Thus, it is reasonable to
try to match the intermediate energy region to QCD. This point is
partly supported by the success of the
chiral perturbation theory~\cite{gl}, which basically is owed to the
existence of a natural small parameter $m_{\pi}^2/m_{\rho}^2\simeq0.03$
in the low-energy strong interactions. On the other hand, the
operator product expansion (OPE) allows to extend the
applicability of the perturbative QCD to the region below
$\Lambda_{\text{CSB}}$, up to the scale $\mu\simeq
m_{\rho}$~\cite{svz}. We arrive thus at the conclusion that the
matching region should be $m_{\rho}<\mu<\Lambda_{\text{CSB}}$.

Let us for a while neglect the pion interactions in the CQM and
consider only almost free constituent quarks sufficiently weakly
interacting by means of low-energy gluons. The $SU(2)$ $V$ and $A$
quark currents can be then simply constructed,
\be
j_{\mu,V}^{\text{CQM}}=\bar{\psi}\gamma_{\mu}\vec{\tau}\psi,\qquad
j_{\mu,A}^{\text{CQM}}=\bar{\psi}\gamma_{\mu}\gamma_5\vec{\tau}\psi.
\ee
Consider the two-point correlators of these currents,
\be
\label{cor2}
\Pi_{\mu\nu,J}^{\text{CQM}}(q^2)=\int d^4x\,e^{-iqx}\left\langle
j_{\mu,J}^{\text{CQM}}(x)j_{\nu,J}^{\text{CQM}}(0)\right\rangle,
\ee
here $J=V,A$.
As long as the effective coupling constant in the CQM is rather
small, one may estimate the V and A correlators~\eqref{cor2} by
doing a standard one-loop perturbative calculation for the polarization
function. Thus,
\begin{gather}
\Pi_{\mu\nu,V}^{\text{CQM}}(q^2)\sim\int d^4\!p\,\text{tr}
\frac{\gamma_{\mu}}{\frac{\not{q}}{2}+\not{\!p}-m_{\text{con}}}
\frac{\gamma_{\nu}}{\frac{\not{q}}{2}-\not{\!p}-m_{\text{con}}},\\
\Pi_{\mu\nu,A}^{\text{CQM}}(q^2)\sim\int d^4\!p\,\text{tr}
\frac{\gamma_{\mu}\gamma_5}{\frac{\not{q}}{2}+\not{\!p}-m_{\text{con}}}
\frac{\gamma_{\nu}\gamma_5}{\frac{\not{q}}{2}-\not{\!p}-m_{\text{con}}},
\end{gather}
where $m_{\text{con}}$ is the constituent quark mass. Taking the
trace and neglecting the irrelevant for us terms quadratic in cut-off,
we obtain,
\begin{gather}
\Pi_{\mu\nu,V}^{\text{CQM}}(q^2)\sim\left\{\left(-\delta_{\mu\nu}q^2+q_{\mu}q_{\nu}\right)F_-+
q_{\mu}q_{\nu}F_+\right\}I(q^2),\\
\Pi_{\mu\nu,A}^{\text{CQM}}(q^2)\sim\left\{\left(-\delta_{\mu\nu}q^2+q_{\mu}q_{\nu}\right)F_++
q_{\mu}q_{\nu}F_-\right\}I(q^2),
\end{gather}
where
\begin{gather}
F_{\pm}=1\pm\frac{4m_{\text{con}}^2}{q^2},\\
I(q^2)=\int d^4\!p\frac{1}{\left(\frac{q}{2}+p\right)^2-m_{\text{con}}^2}
\frac{1}{\left(\frac{q}{2}-p\right)^2-m_{\text{con}}^2}.
\end{gather}
To compare these expressions with the OPE in QCD we have to
perform the Wick rotation and consider the transverse part
$\Pi_{J\bot}^{\text{CQM}}(q^2)$ only,
\begin{gather}
\label{t1}
\Pi_{V\bot}^{\text{CQM}}(Q^2)\sim\left(1-\frac{4m_{\text{con}}^2}{Q^2}\right)I(Q^2),\\
\label{t2}
\Pi_{A\bot}^{\text{CQM}}(Q^2)\sim\left(1+\frac{4m_{\text{con}}^2}{Q^2}\right)I(Q^2).
\end{gather}
Is is seen that the $V$ and $A$ correlators are equal in the limit
of exact chiral symmetry, $m_{\text{con}}\rightarrow0$, and in the
limit of asymptotic chiral symmetry, $Q^2\rightarrow\infty$. In
the limit of vanishing euclidean momentum, $Q^2\rightarrow0$, they
have opposite sign, but equal absolute value. This sign flip could
be regarded as a signal of change of chiral symmetry realization
at low energies. The second term in Eqs.~\eqref{t1} and~\eqref{t2}
emerges due to the chiral symmetry breaking, it is different for
the $V$ and $A$ channels, let us denote it
$\Pi_{\text{CSB},J}^{\text{CQM}}(Q^2)$. Of interest for us is the
fraction
\be
\label{ratio}
\frac{\Pi^{\text{CQM}}_{\text{CSB},V}(Q^2)}{\Pi^{\text{CQM}}_{\text{CSB},A}(Q^2)}=-1.
\ee

The inclusion of pion interactions should correct the simple
picture above because the derivative of pion field enters the
axial-vector current, moreover, in the matching region,
$m_{\rho}<\mu<\Lambda_{\text{CSB}}$, the higher order derivative
terms of the pion field may become significant. However, such
derivative terms can affect strongly the longitudinal parts of the
correlators while they should not couple to the transverse
parts in the chiral limit. Since we are working with the
transverse parts only, it looks reasonable to neglect the terms
involving $\pi$ in Eq.~\eqref{cur}. Moreover, we will match the
CQM to QCD in the large-$N_c$ limit~\cite{hoof}, hence, the same
limit has to be taken from the CQM side, this provides a
suppression of possible multiparticle contributions to
current~\eqref{cur}. Thus, the residual effect of
the strong interactions reduces to the renormalization of the
axial-vector current (factor $g_A$ in Eq.~\eqref{cur}). This
constitutes our first assumption.

Our second assumption concerns a concrete realization of this
renormalization. We propose an alternative interpretation of
the origin of the axial coupling $g_A$ in Eq.~\eqref{cur}.
In QCD, one constructs the $V$ and $A$ currents from the same quark
spinors, but in the effective theory it is not evident that we are
allowed to do this. Actually, the axial-vector sector is affected
strongly by the pion interactions, this may lead to the fact that
we should use another quark spinors in the A-channel, let us
denote this circumstance by a prime. But phenomenological success
of the CQM suggests that neglecting the direct pion contributions,
the action of the $A'$-current may be simulated as the action of
the $A$-current (constructed from the same quark spinors as the
$V$-current) if we accept the following renormalization prescription:
$A=g_AA'$. The $V-A$ current in QCD, Eq.~\eqref{cur2}, turns into
the $V-g_AA'$ current below $\Lambda_{\text{CSB}}$. The identical
notation for the quark spinors in the vector and axial-vector
parts of current~\eqref{cur} should be then understood
symbolically only.

We would provide the following qualitative support in favor of
this hypothesis. If the constituent quarks are almost free, the
corresponding left nucleon current is expected to experience the
same renormalization. The nucleon analogue $j_{\mu,L}^N$ of the left
quark current~\eqref{cur} enters the amplitude of the nucleon $\beta$-decay
and it can be written in the form
\be
j_{\mu,L}^N=\bar{\psi}_p\gamma_{\mu}(1-G_A\gamma_5)\psi_n.
\ee
This current is constructed following the Fermi $V-A$ theory of
weak interactions. Hence, initially one has the $V-A'$ current but
then in calculations one uses the same $\psi_{p,n}$ for the vector
and axial-vector parts, {\it i.e.} one works with the $V-g_A^{-1}A$
hadron current.
Consequently, we expect $g_A\approx G_A^{-1}\approx 0.79$ which is
reasonable (notice that from relation~\eqref{Ga} we would formally
obtain $g_A=\sqrt{0.6}\approx0.77$ which is also not bad). If our
hypothesis is right this numerology is not accidental.

Thus, the assumptions above result in the following conclusion:
In ratio~\eqref{ratio} we had different quark spinors in the
numerator and in the denominator. If we want to compare the
correlators calculated with the same spinors, say with the ones
entering the vector current (let us refer to them as bare spinors)
we should renormalize the $A$-correlator in the following way,
\be
\label{t3}
\Pi^{\text{CQM}}_{\text{CSB},A}(Q^2)\rightarrow
g_A^{-2}\Pi^{\text{CQM}}_{\text{CSB},A}(Q^2).
\ee
This renormalization effectively takes into account the
contribution of pion interactions.

Now we formulate our matching condition between the
CQM and QCD in the region $m_{\rho}<\mu<\Lambda_{\text{CSB}}$,
where the both are expected to describe the same physics related
to CSB. We require that the {\it bare} quark spinors in the CQM Lagrangian
can be replaced by those of the QCD Lagrangian at
$m_{\rho}<\mu<\Lambda_{\text{CSB}}$ with ensuing equality of
operators of quark currents,
\be
\bar{\psi}\gamma_{\mu}\vec{\tau}\psi|_{\text{CQM}}^{\text{bare}}\simeq
\bar{\psi}\gamma_{\mu}\vec{\tau}\psi|_{\text{QCD}},
\ee
\be
\bar{\psi}\gamma_{\mu}\gamma_5\vec{\tau}\psi|_{\text{CQM}}^{\text{bare}}\simeq
\bar{\psi}\gamma_{\mu}\gamma_5\vec{\tau}\psi|_{\text{QCD}}.
\ee
We require also that the same identification is valid between the
fundamental gluon fields in the QCD Lagrangian and the long-range
gluons in the CQM. Then relations~\eqref{ratio} and~\eqref{t3}
lead to
\be
\label{ratio2}
\frac{\Pi^{\text{QCD}}_{\text{CSB},V}(Q^2)}{\Pi^{\text{QCD}}_{\text{CSB},A}(Q^2)}\simeq-g_A^2.
\ee

The problem now is to find $\Pi^{\text{QCD}}_{\text{CSB},J}(Q^2)$, {\it i.e.}
the parts of the corresponding QCD correlators which appear due to the CSB.
A solution of such a task is known in the euclidean region due to
the OPE method~\cite{svz}. Accepting the chiral and large-$N_c$~\cite{hoof}
limits, the OPE for $\Pi_{\mu\nu,J}^{\text{QCD}}(q^2)$ in
the one-loop approximation reads as follows at large euclidean
momentum $Q$,
\ba
\label{V}
\Pi^{\text{QCD}}_{J\bot}(Q^2)&=&\frac{N_c}{12\pi^2}\left(1+\frac{\alpha_s}{\pi}\right)
\ln\!\frac{\mu^2}{Q^2}+\frac{\alpha_s}{12\pi}\frac{\langle G^2\rangle}{Q^4} \nl
&&+\frac{4\pi\xi_J\alpha_s}{9}
\frac{\langle\bar{q}q\rangle^2}{Q^6}+\mathcal{O}\left(\frac{1}{Q^8}\right),
\ea
where
\be
\label{ksi}
\xi_V=-7,\qquad \xi_A=11,
\ee
and we have defined
\be
\label{trans}
\Pi_{\mu\nu,J}^{\text{QCD}}(Q^2)=\left(-\delta_{\mu\nu}Q^2+Q_{\mu}Q_{\nu}\right)
\Pi^{\text{QCD}}_{J\bot}(Q^2).
\ee
The symbols $\langle G^2\rangle$ and $\langle\bar{q}q\rangle$ denote the
gluon and quark condensate, respectively. The power-like expansion~\eqref{V} shows
explicitly that the CSB effects set in since the $\mathcal{O}(1/Q^6)$
terms. This agrees with our naive model calculation above --- expanding Eqs.~\eqref{t1}
and~\eqref{t2} at large $Q^2$ we
obtain the same qualitative behaviour for the CSB contribution.
The part of $\Pi^{\text{QCD}}_{J\bot}(Q^2)$ which absorbs the
leading contributions related to the CSB (more precisely, the contributions
which are different for the $V$ and $A$ channels),
$\Pi^{\text{QCD}}_{\text{CSB},J}(Q^2)$, is evident from Eq.~\eqref{V},
\be
\label{ansn}
\Pi^{\text{QCD}}_{\text{CSB},J}(Q^2)=
\frac{4\pi\xi_J\alpha_s}{9}
\frac{\langle\bar{q}q\rangle^2}{Q^6}+\mathcal{O}\left(\frac{1}{Q^8}\right).
\ee
The relation above gives
\be
\label{ratio3}
\frac{\Pi^{\text{QCD}}_{\text{CSB},V}(Q^2)}{\Pi^{\text{QCD}}_{\text{CSB},A}(Q^2)}=
\frac{\xi_V}{\xi_A}+\mathcal{O}\left(\frac{1}{Q^2}\right).
\ee
Collecting Eqs.~\eqref{ratio2},~\eqref{ratio3}, and~\eqref{ksi},
we get our final result,
\be
g_A^2\simeq\frac{7}{11},
\ee
which yields $g_A\approx0.80$ in a good agreement with the existing
phenomenological estimates.

The obtained value calls for a comment concerning the large-$N_c$
behaviour of $g_A$. In the literature there is a discrepancy with
regard to the question of whether or not $g_A=1$ in the
large-$N_c$ limit, see~\cite{rafael,bron,jacob} for discussions. Our estimate
has been performed taking this limit from the outset since it was
used in OPE~\eqref{V} --- the factorized form of the numerator in
$\mathcal{O}(1/Q^6)$ term takes place by virtue of the vacuum
saturation hypothesis, which is justified in the large-$N_c$ limit
only~\cite{svz}. We conclude thus that the deviation of $g_A$ from
unity is not an artefact of the $\mathcal{O}(1/N_c)$ corrections,
at least within the presented approach.

Let us summarize our scheme. We have considered the transverse
parts of vector and axial-vector two-point correlators and
extracted the leading contributions coming from the spontaneous
chiral symmetry breaking, which are different for the vector
and axial-vector channels. In the constituent quark model, the
ratio of these contributions in the vector and axial-vector
channels is $-1$ whereas from the QCD side in the large-$N_c$
limit it is equal to $-7/11$. We assumed that the difference appears
mainly from the
fact that the QCD vector and axial-vector currents, $j_{\mu}^V$
and $j_{\mu}^A$, are built from the same quark spinors while in
the constituent quark model this is not the case, the effect can
be effectively described as the renormalization of the current $j_{\mu}^A$,
$j_{\mu}^A|_{\text{ren}}=g_A^{-1}j_{\mu}^A|_{\text{bare}}$, and, in fact,
by default one uses the renormalized current in the two-point
correlator $\langle j_{\mu}^Aj_{\nu}^A\rangle$. The correct
matching with QCD correlators, however, should be achieved
with the unrenormalized currents if we want to escape the double counting of
non-perturbative effects. Thus, the axial-vector correlator of the
constituent quark model should be multiplied by the factor
$g_A^{-2}$ when doing matching to QCD. This gives immediately the
relation for the axial coupling, $g_A^2\simeq7/11$.

Finally, realizing that the undertaking reasoning is expected, at
best, to give an order-by-magnitude estimate, it looks quite
spectacular that the obtained value for $g_A$ is so close to the
accepted phenomenological estimate. Perhaps, this somewhat
justifies {\it a posteriori} the made assumptions.

A discussion of the work with A. A. Andrianov and D. Espriu is gratefully
acknowledged. The work was supported by the Ministry of Education of Russian
Federation, grant RNP.2.1.1.1112.

\end{document}